\documentclass[journal]{IEEEtran}
\usepackage{amsmath}
\usepackage{algorithm}
\usepackage{algorithmic}
\usepackage{subfigure}
\usepackage{cite}
\usepackage{makecell}
\usepackage{url}
\usepackage{slashbox}
\usepackage[breaklinks]{hyperref}
\usepackage{epsfig,psfig}
\usepackage{graphicx}
\usepackage{dblfloatfix}
\usepackage{caption}
\usepackage[table]{xcolor}
\usepackage{array}

\usepackage{tabularx}
\usepackage{colortbl}
\usepackage{pifont}
\usepackage{arydshln}
\usepackage{subcaption}

\begin{document}

\title{SESO-ISAC: Service-Aware End-to-End Sensing Orchestration for 6G ISAC}

\author{Youbin Jeon, Laeyoung Kim, Myungjune Youn, and Sangheon Pack,~\IEEEmembership{Senior Member,~IEEE}
\IEEEcompsocitemizethanks{
\IEEEcompsocthanksitem Y. Jeon, L. Kim, and M. Youn are with LG Electronics, Seoul, South Korea. E-mail: \{youbin.jeon, laeyoung.kim, m.youn\}@lge.com. S. Pack is with the School of Electrical Engineering, Korea University, Seoul 02841, South Korea. E-mail: shpack@korea.ac.kr}%
\thanks{(Corresponding author: Youbin Jeon and Sangheon Pack.)}}
\maketitle

\begin{abstract}
Unlike conventional communication services, 6G Integrated Sensing and Communication (ISAC) services require complete end-to-end sensing operations, from data collection to processing, result generation, and exposure. In this article, we propose SESO-ISAC, a service-aware end-to-end sensing orchestration framework that maps a sensing service request to a complete sensing configuration under the sensing execution context. Rather than treating sensing configuration decisions independently, SESO-ISAC captures their interdependence across the end-to-end sensing operation, evaluates candidate configurations according to the applicable QoS requirements, and re-evaluates the selected configuration as the sensing execution context changes. Through two representative case studies, we show that the preferred sensing configuration depends on the overall performance across the applicable QoS dimensions and can change during operation.
\end{abstract}


\IEEEpeerreviewmaketitle

\section{Introduction}

Integrated Sensing and Communication (ISAC) is considered a key technology for 6G. It extends the role of mobile networks from data delivery to environmental awareness by integrating sensing capability into communication systems~\cite{Xuewen25}. The Third Generation Partnership Project (3GPP) first defined ISAC use cases and service requirements in Release 19~\cite{TS22.137}. In Release 20, 3GPP SA WG2 has been studying the architecture and end-to-end procedures to support ISAC~\cite{Lin26}. The 5G-Advanced ISAC architecture mainly focuses on aerial-object detection and tracking based on gNB-based sensing~\cite{TS23.137}. In contrast, the 6G ISAC architecture study considers broader sensing scenarios, including the participation of user equipment (UE) as well as radio access network (RAN) nodes in sensing operations~\cite{TR23.801-01}. This broader architectural scope is motivated by diverse 6G ISAC use cases, such as industrial automation, vulnerable pedestrian protection, and environmental monitoring, which are often associated with safety-critical or mission-critical services~\cite{Lee26,Nabati25}. For such services, the network needs to provide sensing results with the required latency, reliability, freshness, and sensing quality.

In a conventional communication service, the network focuses on the connectivity configuration and path setup to deliver data from a source to a destination~\cite{TS23.501}. In contrast, an ISAC sensing service needs to perform sensing execution and processing before delivering the sensing results to the sensing service consumer (SSC). A sensing service request specifies the sensing service type and quality of service (QoS) requirements requested by the SSC. The network then needs to map the request to an executable end-to-end sensing configuration by considering the available sensing entities (SenEs), their capabilities, and the available processing and reporting paths.

In the literature, existing work has addressed individual aspects of this request-to-configuration mapping problem. Existing ISAC studies have mainly focused on radio-level problems, such as waveform design, beamforming, radio resource allocation, and sensing accuracy~\cite{Liu26,Ge25,Li25}. Recent studies have also investigated service-aware resource allocation and sensing-topology switching with processing architectures~\cite{Dong24,Lyazidi25}. These studies consider only a subset of the decisions required to form an end-to-end sensing configuration. Meanwhile, 3GPP SA2 has been discussing functional entities and end-to-end procedures for ISAC. However, existing work has not sufficiently addressed how interdependent sensing decisions can be composed into complete and executable end-to-end sensing operations. This mapping is particularly challenging because individual configuration decisions are tightly coupled. Therefore, satisfying diverse service requirements requires holistic and fine-grained sensing orchestration that jointly determines SenE participation, Tx/Rx roles, sensing mode, processing point, and reporting/exposure path across the entire sensing lifecycle.


\definecolor{ProfileHeaderBG}{RGB}{235,231,222}
\definecolor{ProfileGrid}{RGB}{176,170,160}

\newcolumntype{C}[1]{%
    >{\centering\arraybackslash}m{#1}%
}
\newcolumntype{L}[1]{%
    >{\raggedright\arraybackslash}m{#1}%
}

\newcommand{\tblitem}[1]{%
    \par\noindent
    \hangindent=1.15em
    \hangafter=1
    \(\triangleright\)\hspace{0.4em}#1%
}

\newcommand{\headercell}[1]{%
    \cellcolor{ProfileHeaderBG}%
    \rule[-1.5ex]{0pt}{5.5ex}%
    \bfseries\makecell[c]{#1}%
}

\begin{table*}[t]

\caption{Sensing requirement profiles for 3GPP SA1 ISAC use cases.}
\label{tab:tab_1}

\centering
\begin{minipage}{0.97\textwidth}
\centering

\begingroup
\footnotesize


\setlength{\tabcolsep}{2.8pt}
\setlength{\parindent}{0pt}
\setlength{\parskip}{0pt}

\setlength{\arrayrulewidth}{0.6pt}
\arrayrulecolor{ProfileGrid}

\renewcommand{\arraystretch}{2.5}
\setlength{\extrarowheight}{0pt}

\begin{tabular}{
    |C{0.16\textwidth}
    |L{0.33\textwidth}
    |L{0.20\textwidth}
    |L{0.25\textwidth}|
}
\hline


\headercell{Sensing profile} &
\headercell{Representative use cases} &
\headercell{Characteristic \\QoS dimension} &
\headercell{Defining criterion} \\
\hline


\bfseries
\makecell{Latency-critical\\interaction} &
\tblitem{AMR collision avoidance in smart factories}
\tblitem{Enhanced XR user navigation}
\tblitem{Safety assistance for vulnerable pedestrians} &
\tblitem{Sensing latency} &
\tblitem{Result availability within the service deadline} \\
\hline


\bfseries
\makecell{Reliability-critical\\detection and tracking} &
\tblitem{UAV flight trajectory tracking}
\tblitem{AGV detection and tracking in factories}
\tblitem{UAV intrusion detection} &
\tblitem{Sensing reliability} &
\tblitem{Missed-detection and false-alarm requirements} \\
\hline


\bfseries
\makecell{Freshness-sensitive\\continuous sensing} &
\tblitem{Real-time traffic flow monitoring}
\tblitem{Rainfall and flooding monitoring}
\tblitem{Health and sports monitoring} &
\tblitem{Sensing freshness} &
\tblitem{Sensing-result updates at the service-required refresh interval} \\
\hline


\bfseries
\makecell{Quality-intensive\\sensing} &
\tblitem{High-resolution topographical mapping}
\tblitem{Environmental object reconstruction}
\tblitem{Gesture recognition in industrial environments} &
\tblitem{Sensing quality} &
\tblitem{Task-usable sensing accuracy and resolution} \\
\hline

\end{tabular}

\endgroup
\end{minipage}

\end{table*}

To address this problem, we propose SESO-ISAC, a service-aware end-to-end sensing orchestration framework for ISAC. The main contributions are summarized as follows: 1) SESO-ISAC defines an end-to-end sensing lifecycle and jointly orchestrates SenE participation, Tx/Rx role assignment, sensing mode, processing point, and reporting/exposure path to form a complete end-to-end configuration; 2) our case studies illustrate a multi-dimensional configuration selection method that filters candidates against mandatory QoS boundaries and compares the feasible candidates across applicable QoS dimensions; and 3) SESO-ISAC supports re-evaluation in response to structural and performance changes, enabling the end-to-end sensing operation to be retained or reselected as the execution context changes.

The remainder of this article first introduces 6G ISAC service models, including sensing service profiles, sensing entities, modes, and configurations. It then presents the SESO-ISAC framework and applies it to two representative sensing services. Finally, we discuss open research issues and conclude the article.

\section{6G ISAC Service Models}
\label{Sec:backandmotiv}
In this section, we describe 1) sensing service profiles and 2) sensing entities, modes, and configurations.

\subsection{Sensing Service Profiles}
3GPP SA1 ISAC use cases exhibit heterogeneous sensing requirements~\cite{TS22.137,TR22.870}. From a service-level perspective, these requirements can be grouped according to the QoS dimension that most directly affects the suitability of a sensing result for its intended service objective. For time-critical interaction services, delayed sensing results can reduce their usefulness for the intended service action, making sensing latency a key requirement. For detection and tracking services, service continuity depends on reliable target observation over time, making missed-detection and false-alarm performance the principal indicators of sensing reliability. For continuous monitoring services, service usability depends on sensing-result updates at a rate sufficient to reflect the current state of the monitored target or environment, making the service-required refresh interval the primary freshness criterion. For sensing tasks requiring detailed characterization of a target or environment, service usability depends on the fidelity of the generated sensing result, making sensing accuracy and resolution the defining quality requirements.

Table~\ref{tab:tab_1} summarizes representative sensing requirement profiles based on these characteristic QoS dimensions. These profiles are not intended to be an exhaustive taxonomy of all 3GPP SA1 and 6G ISAC use cases, nor do they prescribe the complete set of QoS requirements for a sensing request. Instead, they provide a service-level abstraction of representative sensing requirement characteristics.

\begin{figure*}
\centering
\includegraphics[width=7in]{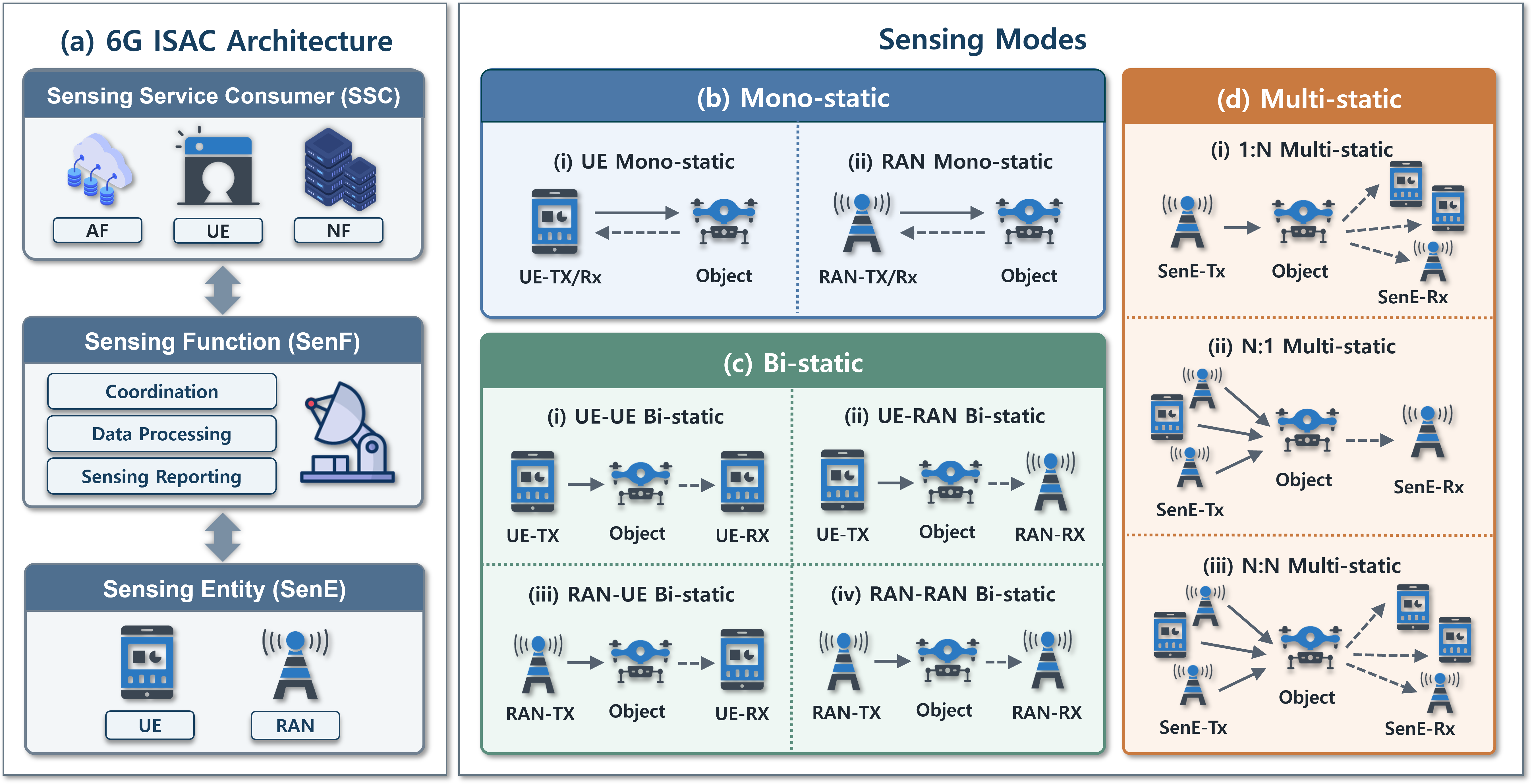}
\caption{6G ISAC architecture and sensing modes: (a) 6G ISAC architecture; (b) mono-static sensing; (c) bi-static sensing; (d) multi-static sensing.}
\label{fig:fig_1}
\end{figure*}

\subsection{Sensing Entities, Modes and Configurations}
As shown in Fig.~\ref{fig:fig_1}(a), from a 3GPP SA2 perspective, an ISAC sensing service is supported through interactions among SSC, sensing function (SenF), and SenEs. The SSC requests a sensing service and consumes the provided sensing result. The SSC can be a UE, an application function (AF), or a network function (NF). In the current 3GPP SA2 work, the SenF is considered a core network function that handles the sensing service request and coordinates the functions involved in sensing operation configuration, sensing-data processing, result reporting, and exposure. A SenE is an entity that transmits or receives sensing signals, which can be a UE or a RAN node. Each SenE may support a sensing transmitter (Tx) role, a sensing receiver (Rx) role, or both roles. The sensing role is assigned per sensing operation. Even for the same sensing service type, different SenE combinations and Tx/Rx role assignments may be selected depending on the SSC type, sensing target or area, sensing requirements, and the capabilities and locations of available SenEs.

Figure~\ref{fig:fig_1}(b)--(d) illustrate mono-static, bi-static, and multi-static sensing modes. In mono-static sensing (see Fig.~\ref{fig:fig_1}(b)), one SenE performs both Tx and Rx roles. This reduces inter-SenE coordination and sensing-data transfer, while the achievable sensing performance depends on the capability and location of that SenE. In bi-static sensing (see Fig.~\ref{fig:fig_1}(c)), different SenEs perform the Tx and Rx roles. Thus, UE--UE, UE--RAN, RAN--UE, and RAN--RAN configurations can be considered, where the first and second terms denote the SenE-Tx and the SenE-Rx. Separating the Tx and Rx locations can improve target observability under suitable sensing geometry but requires inter-SenE coordination. When sensing-data processing is performed outside the receiving SenE, the collected sensing data also need to be transferred to the selected processing point. In multi-static sensing (see Fig.~\ref{fig:fig_1}(d)), multiple SenEs participate as Tx and/or Rx in a sensing operation. The 1:N, N:1, and N:N configurations can provide multiple observations of the same sensing target or area, but require coordination among participating SenEs as well as sensing-data collection and association. As the number of participating SenEs increases, the selection of the processing point and the associated sensing-data transfer paths becomes increasingly important.

\begin{figure*}
\centering
\includegraphics[width=7in]{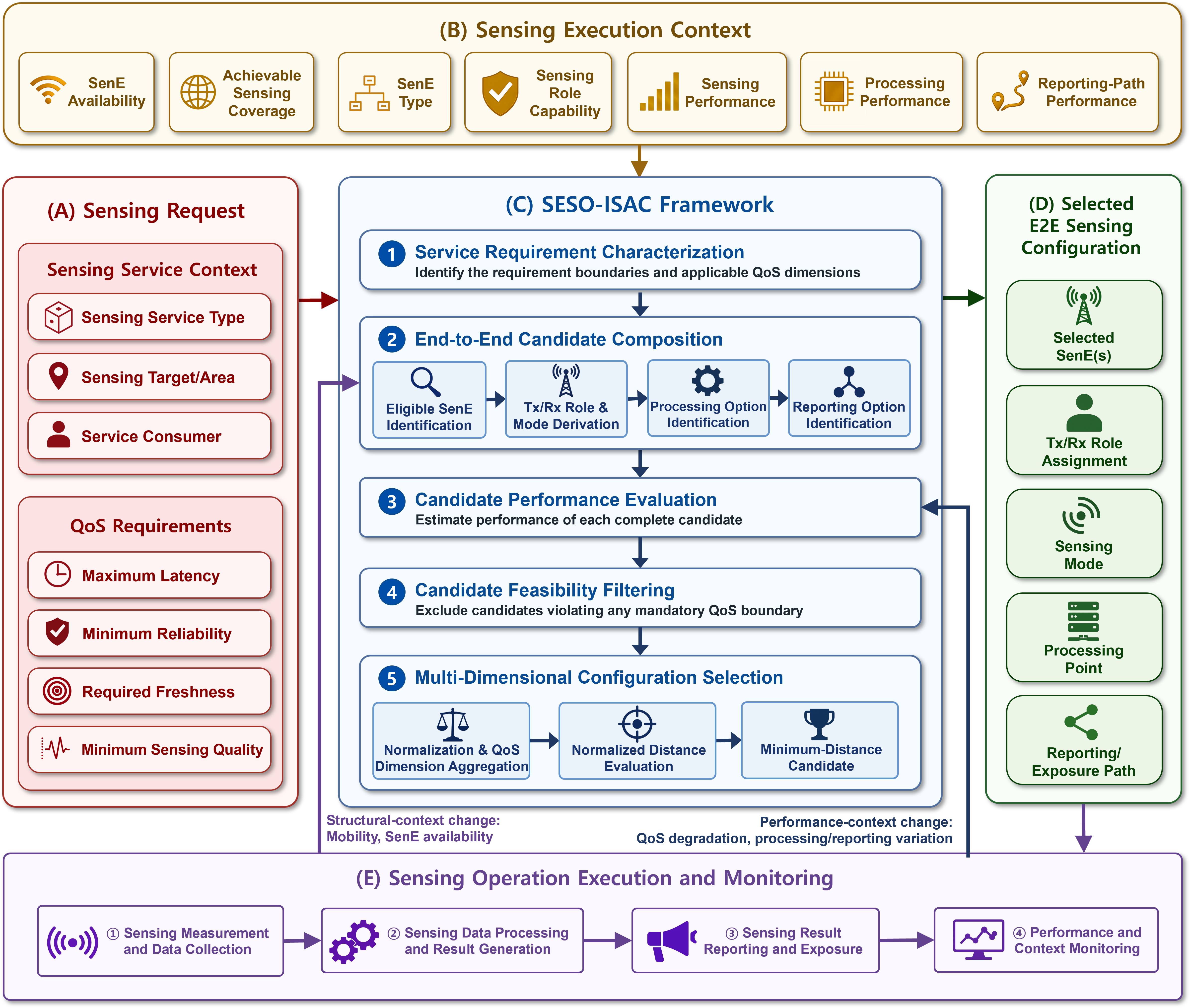}
\caption{SESO-ISAC Orchestration.}
\label{fig:fig_4}
\end{figure*}

\section{SESO-ISAC Framework}
\label{Sec:usecases}
This section presents SESO-ISAC for composing, evaluating, selecting, and re-evaluating complete end-to-end sensing configurations.

\subsection{Sensing Lifecycle and Orchestration}
Figure~\ref{fig:fig_4} shows the end-to-end sensing lifecycle. The SSC sends a sensing request that specifies the sensing service type, sensing target or area, and QoS requirements (see (A) in Fig.~\ref{fig:fig_4}). The SESO-ISAC framework selects an end-to-end sensing configuration considering the sensing execution context (see (B)--(C) in Fig.~\ref{fig:fig_4}). Based on the selected configuration, the network configures the participating SenEs, Tx/Rx roles, sensing mode, processing point, and reporting/exposure path (see (D) in Fig.~\ref{fig:fig_4}). The selected SenEs perform sensing measurements and collect sensing data according to the assigned Tx/Rx roles (see (E)-\textcircled{\raisebox{-1.0pt}{1}} in Fig.~\ref{fig:fig_4}). The collected data are processed at the configured processing point to generate a sensing result (see (E)-\textcircled{\raisebox{-1.0pt}{2}} in Fig.~\ref{fig:fig_4}). After that, the sensing result is delivered to the SSC through the configured reporting/exposure path (see (E)-\textcircled{\raisebox{-1.0pt}{3}} in Fig.~\ref{fig:fig_4}).

For periodic or event-triggered services, sensing execution and result reporting are repeated. During repeated sensing operations, the sensing execution context can change. Specifically, the structural context captures the sensing geometry and participating SenEs, which can be affected by mobility and SenE availability. The performance context captures processing and reporting conditions that can affect the performance of the current configuration. Therefore, SESO-ISAC monitors the sensing performance and execution context (see (E)-\textcircled{\raisebox{-1.0pt}{4}} in Fig.~\ref{fig:fig_4}) to determine whether a configuration re-evaluation is required.



\definecolor{TIIHeaderOneBG}{RGB}{232,226,242}
\definecolor{TIIHeaderTwoBG}{RGB}{225,236,248}

\definecolor{TIIGridOne}{RGB}{178,168,198}
\definecolor{TIIGridTwo}{RGB}{156,182,212}

\definecolor{TIIBadBG}{RGB}{252,238,238}


\newcommand{\TIIcmark}{%
    \textcolor{green!55!black}{\ding{51}}%
}

\newcommand{\TIIxmark}{%
    \textcolor{red!75!black}{\ding{55}}%
}

\newcommand{\TIIvio}[1]{%
    \cellcolor{TIIBadBG}\textbf{#1}%
}

\newcommand{\TIIbest}[1]{%
    \textbf{#1}%
}

\newcommand{\TIIreqstrut}{%
    \rule[-1.2ex]{0pt}{5.2ex}%
}


\begin{table*}[t]
\centering

\caption{Performance comparison and configuration selection for
(a) factory robot collision avoidance and
(b) UAV flight trajectory tracking.}
\label{tab:case-results}

\begingroup

\renewcommand{\arraystretch}{1.35}
\setlength{\tabcolsep}{3.5pt}


{\small
\arrayrulecolor{TIIGridOne}

\begin{tabular}
{|>{\centering\arraybackslash}m{1.90cm}|
 *{8}{>{\centering\arraybackslash}m{1.70cm}|}}
\hline


\cellcolor{TIIHeaderOneBG}
\makecell[c]{\textbf{Configuration}}
&
\cellcolor{TIIHeaderOneBG}
\makecell[c]{\textbf{Sensing}\\
             \textbf{latency}\\
             \textbf{[ms]}}
&
\cellcolor{TIIHeaderOneBG}
\makecell[c]{\textbf{Positioning}\\
             \textbf{error [m]}}
&
\cellcolor{TIIHeaderOneBG}
\makecell[c]{\textbf{Velocity}\\
             \textbf{error [m/s]}}
&
\cellcolor{TIIHeaderOneBG}
\makecell[c]{\textbf{Range}\\
             \textbf{resolution}\\
             \textbf{[m]}}
&
\cellcolor{TIIHeaderOneBG}
\makecell[c]{\textbf{False alarm}\\
             \textbf{[\%]}}
&
\cellcolor{TIIHeaderOneBG}
\makecell[c]{\textbf{Refresh}\\
             \textbf{interval}\\
             \textbf{[s]}}
&
\cellcolor{TIIHeaderOneBG}
\makecell[c]{\textbf{Feasible}}
&
\cellcolor{TIIHeaderOneBG}
\makecell[c]{\textbf{Normalized}\\
             \textbf{distance}}
\\
\hline


\TIIreqstrut
\makecell[c]{\textbf{3GPP}\\
             \textbf{requirement}}
&
\(< 500\)
&
\(\leq 1.0\)
&
\(\leq 1.0\)
&
\(\leq 1.0\)
&
\(\leq 5.0\)
&
\(\leq 0.05\)
&
--
&
--
\\
\hline


\textbf{\(RC_1\)}
&
80
&
\TIIvio{1.20}
&
\TIIvio{1.10}
&
\TIIvio{1.15}
&
4.0
&
0.050
&
\TIIxmark~No
&
--
\\
\hline


\textbf{\(RC_2\)}
&
150
&
0.75
&
0.80
&
0.80
&
4.0
&
0.045
&
\TIIcmark~Yes
&
0.734
\\
\hline


\textbf{\(RC_3\)}
&
230
&
0.45
&
0.55
&
0.55
&
2.5
&
0.0265
&
\TIIcmark~Yes
&
\TIIbest{0.503}
\\
\hline


\textbf{\(RC_4\)}
&
\TIIvio{520}
&
0.20
&
0.20
&
0.25
&
1.0
&
0.020
&
\TIIxmark~No
&
--
\\
\hline

\end{tabular}
}

\vspace{0.55em}

{\small (a)}

\vspace{1.25em}


{\small
\arrayrulecolor{TIIGridTwo}

\begin{tabular}
{|>{\centering\arraybackslash}m{1.90cm}|
 *{8}{>{\centering\arraybackslash}m{1.70cm}|}}
\hline


\cellcolor{TIIHeaderTwoBG}
\makecell[c]{\textbf{Configuration}}
&
\cellcolor{TIIHeaderTwoBG}
\makecell[c]{\textbf{Sensing}\\
             \textbf{latency}\\
             \textbf{[ms]}}
&
\cellcolor{TIIHeaderTwoBG}
\makecell[c]{\textbf{Positioning}\\
             \textbf{error [m]}}
&
\cellcolor{TIIHeaderTwoBG}
\makecell[c]{\textbf{Velocity}\\
             \textbf{error [m/s]}}
&
\cellcolor{TIIHeaderTwoBG}
\makecell[c]{\textbf{Missed}\\
             \textbf{detection}\\
             \textbf{[\%]}}
&
\cellcolor{TIIHeaderTwoBG}
\makecell[c]{\textbf{False alarm}\\
             \textbf{[\%]}}
&
\cellcolor{TIIHeaderTwoBG}
\makecell[c]{\textbf{Refresh}\\
             \textbf{interval}\\
             \textbf{[s]}}
&
\cellcolor{TIIHeaderTwoBG}
\makecell[c]{\textbf{Feasible}}
&
\cellcolor{TIIHeaderTwoBG}
\makecell[c]{\textbf{Normalized}\\
             \textbf{distance}}
\\
\hline


\TIIreqstrut
\makecell[c]{\textbf{3GPP}\\
             \textbf{requirement}}
&
\(\leq 1000\)
&
\(\leq 2.0\)
&
\(\leq 2.0\)
&
\(\leq 5.0\)
&
\(\leq 5.0\)
&
\(\leq 1.0\)
&
--
&
--
\\
\hline


\textbf{\(UC_1\) (\(t_0\))}
&
150
&
\TIIvio{2.50}
&
\TIIvio{2.20}
&
\TIIvio{6.0}
&
4.0
&
\TIIvio{1.10}
&
\TIIxmark~No
&
--
\\
\hline


\textbf{\(UC_2\) (\(t_0\))}
&
220
&
1.00
&
1.10
&
2.5
&
2.75
&
0.55
&
\TIIcmark~Yes
&
\TIIbest{0.475}
\\
\hline


\textbf{\(UC_3\) (\(t_0\))}
&
900
&
0.50
&
0.60
&
1.25
&
1.50
&
0.25
&
\TIIcmark~Yes
&
0.506
\\
\hline


\textbf{\(UC_1\) (\(t_1\))}
&
170
&
\TIIvio{2.80}
&
\TIIvio{2.60}
&
\TIIvio{7.0}
&
4.5
&
\TIIvio{1.20}
&
\TIIxmark~No
&
--
\\
\hline


\textbf{\(UC_2\) (\(t_1\))}
&
300
&
1.80
&
1.70
&
4.5
&
4.0
&
1.00
&
\TIIcmark~Yes
&
0.803
\\
\hline


\textbf{\(UC_3\) (\(t_1\))}
&
900
&
0.40
&
0.40
&
1.0
&
1.0
&
0.26
&
\TIIcmark~Yes
&
\TIIbest{0.489}
\\
\hline

\end{tabular}
}

\vspace{0.55em}

{\small (b)}

\endgroup

\end{table*}

\subsection{SESO-ISAC Procedure}
Figure~\ref{fig:fig_4}(C) shows the SESO-ISAC framework for determining an end-to-end sensing configuration based on the sensing request and sensing execution context. In this article, the SenF is assumed to coordinate the configuration decision. SESO-ISAC is designed around three key considerations: interdependence among end-to-end configuration elements, multi-dimensional QoS evaluation, and configuration re-evaluation under a changing sensing execution context. To address these considerations, SESO-ISAC consists of five steps, as shown in Fig.~\ref{fig:fig_4}(C): 1) service requirement characterization, 2) end-to-end candidate composition, 3) candidate performance evaluation, 4) candidate feasibility filtering, and 5) multi-dimensional configuration selection.

In the first step (see (C)-\textcircled{\raisebox{-1.0pt}{1}} in Fig.~\ref{fig:fig_4}), SESO-ISAC characterizes the sensing request to identify the applicable QoS dimensions and their requirement boundaries. The applicable dimensions are selected from latency, reliability, freshness, and sensing quality. The set of applicable dimensions varies across sensing service requests according to the QoS requirements specified in each request. Each dimension is represented by one or more service-specific metrics, whose specified requirements define the mandatory QoS boundaries used for subsequent candidate evaluation.

In the second step (see (C)-\textcircled{\raisebox{-1.0pt}{2}} in Fig.~\ref{fig:fig_4}), SESO-ISAC composes end-to-end candidate configurations based on the sensing service context and sensing execution context. SESO-ISAC identifies available SenEs and derives possible Tx/Rx role assignments and sensing modes based on the SenE types and sensing role capabilities. Each SenE–role–mode combination determines where sensing measurements are performed and where sensing data are collected, thereby constraining the available processing options. SESO-ISAC identifies available processing options based on their processing performance. Available reporting/exposure paths are determined based on the processing point, SSC location, and path availability. These dependencies are considered jointly so that each candidate forms a complete end-to-end configuration consisting of SenE participation, Tx/Rx role assignment, sensing mode, processing point, and reporting/exposure path.

In the third step (see (C)-\textcircled{\raisebox{-1.0pt}{3}} in Fig.~\ref{fig:fig_4}), SESO-ISAC evaluates the performance of each candidate under the sensing execution context. Each candidate is characterized by the applicable elements of $\mathbf{p}=[L,\mathbf{R},F,\mathbf{Q}]$, where $L$ and $F$ denote sensing latency and freshness, respectively, while $\mathbf{R}$ and $\mathbf{Q}$ contain the applicable reliability metrics (e.g., missed-detection and false-alarm probabilities) and sensing-quality metrics (e.g., positioning/velocity accuracy and sensing resolution). The candidate performance can vary as the execution context changes over time. For example, the mobility of a SenE or sensing target can alter sensing geometry, while changes in processing or reporting conditions can affect latency, freshness, reliability, or quality.

In the fourth step (see (C)-\textcircled{\raisebox{-1.0pt}{4}} in Fig.~\ref{fig:fig_4}), SESO-ISAC filters the candidates against the mandatory QoS boundaries. A candidate that violates any mandatory QoS boundary is excluded because superior performance in other dimensions cannot compensate for the violation. Candidates satisfying all mandatory QoS boundaries proceed to the final selection step. If no feasible candidate remains, the sensing service request cannot be satisfied under the sensing execution context.

In the fifth step (see (C)-\textcircled{\raisebox{-1.0pt}{5}} in Fig.~\ref{fig:fig_4}), SESO-ISAC compares the remaining feasible candidates across the applicable QoS dimensions. Each performance metric is normalized with respect to its mandatory QoS boundary, such that the boundary is mapped to one and better performance approaches zero. When a QoS dimension contains multiple metrics, their normalized values are aggregated using the root mean square (RMS) to obtain a single dimension score. SESO-ISAC then computes the Euclidean norm of the applicable QoS-dimension scores and normalizes it by the square root of the number of applicable dimensions. The candidate with the minimum normalized distance is selected.

The final selection determines the end-to-end sensing configuration (see (D) in Fig.~\ref{fig:fig_4}), which is then used for executing the sensing operation (see (E) in Fig.~\ref{fig:fig_4}). During operation, SESO-ISAC monitors sensing performance and the execution context to determine whether re-evaluation is required. If a structural-context change affects candidate composition, SESO-ISAC repeats the procedure from the candidate composition step (see (C)-\textcircled{\raisebox{-1.0pt}{2}} in Fig.~\ref{fig:fig_4}). If a performance-context change affects candidate performance, SESO-ISAC repeats the procedure from the performance evaluation step (see (C)-\textcircled{\raisebox{-1.0pt}{3}} in Fig.~\ref{fig:fig_4}). The current configuration is retained when it remains feasible and preferred. Otherwise, SESO-ISAC selects a new configuration.

\section{Case Studies}
\label{Sec:performance}
In this section, we apply SESO-ISAC to two representative ISAC use cases: 1) factory robot collision avoidance and 2) UAV flight trajectory tracking. For these case studies, the mandatory QoS boundaries are derived from 3GPP sensing KPIs~\cite{TR22.837}. Representative candidate performance values are used to illustrate the SESO-ISAC procedure. The first case illustrates multi-dimensional configuration selection under a given execution context, while the second case illustrates configuration re-evaluation in response to mobility.

\subsection{Case 1: factory robot collision avoidance}
We consider a continuous sensing service in which a moving autonomous mobile robot (AMR) acts as the SSC and detects workers or obstacles to avoid collisions. In this use case, the limited sensing range of a single AMR and the blockage caused by factory equipment are identified as sensing challenges~\cite{TR22.837}. SESO-ISAC considers four candidate configurations: 1) $RC_1$ uses UE mono-static sensing with local processing; 2) $RC_2$ uses RAN-UE bi-static sensing with UE-local processing; 3) $RC_3$ uses the same sensing mode as $RC_2$ with network-side processing; and 4) $RC_4$ uses observations from multiple SenEs with network-side processing. For result delivery, $RC_1$ and $RC_2$ use local delivery, while $RC_3$ and $RC_4$ use network-to-UE delivery.

\begin{figure*}[t]
    \centering
    \subfigure[]{
        \includegraphics[
            width=0.31\textwidth
        ]{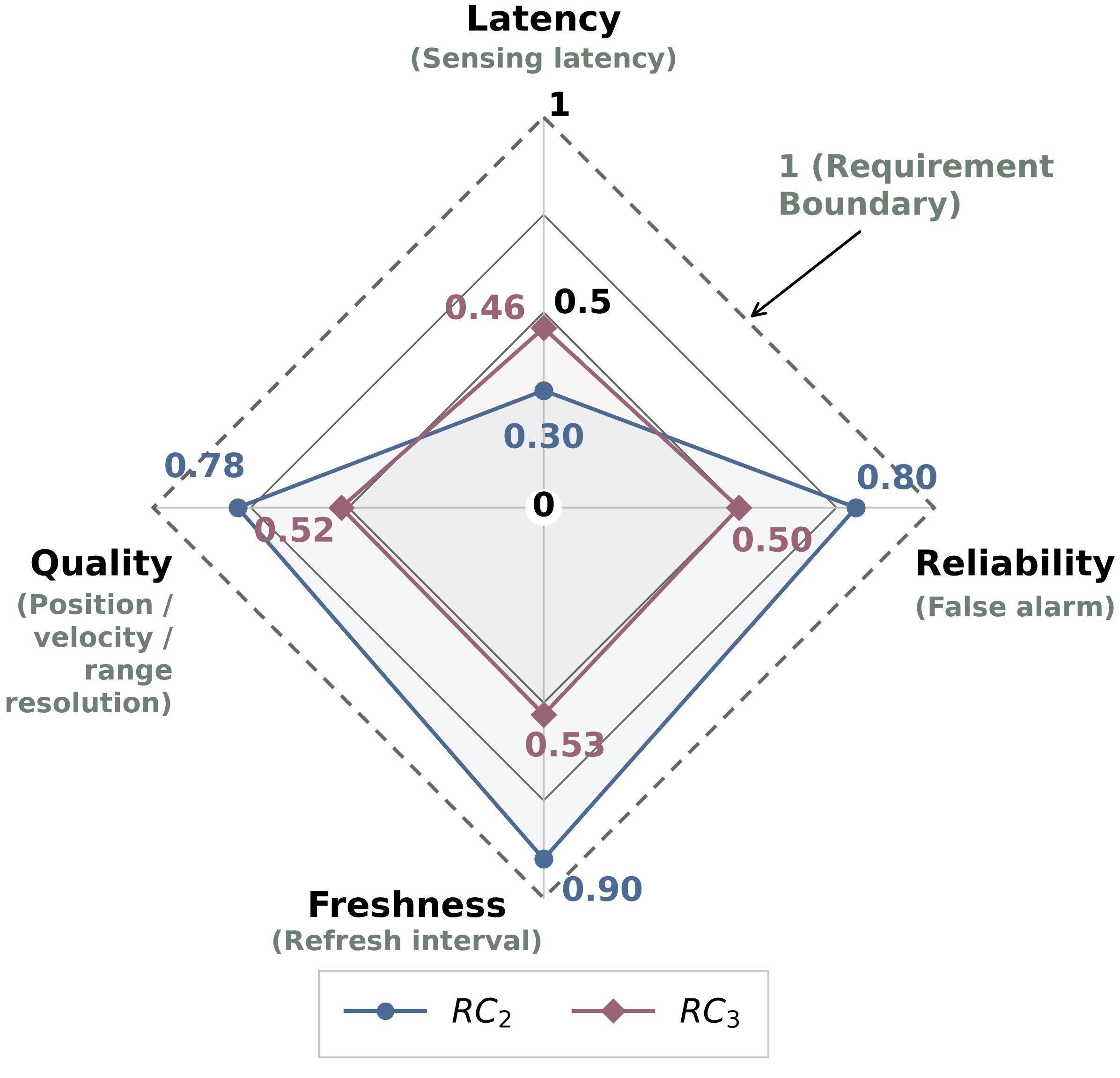}
    }
    \hfill
    \subfigure[]{
        \includegraphics[
            width=0.31\textwidth
        ]{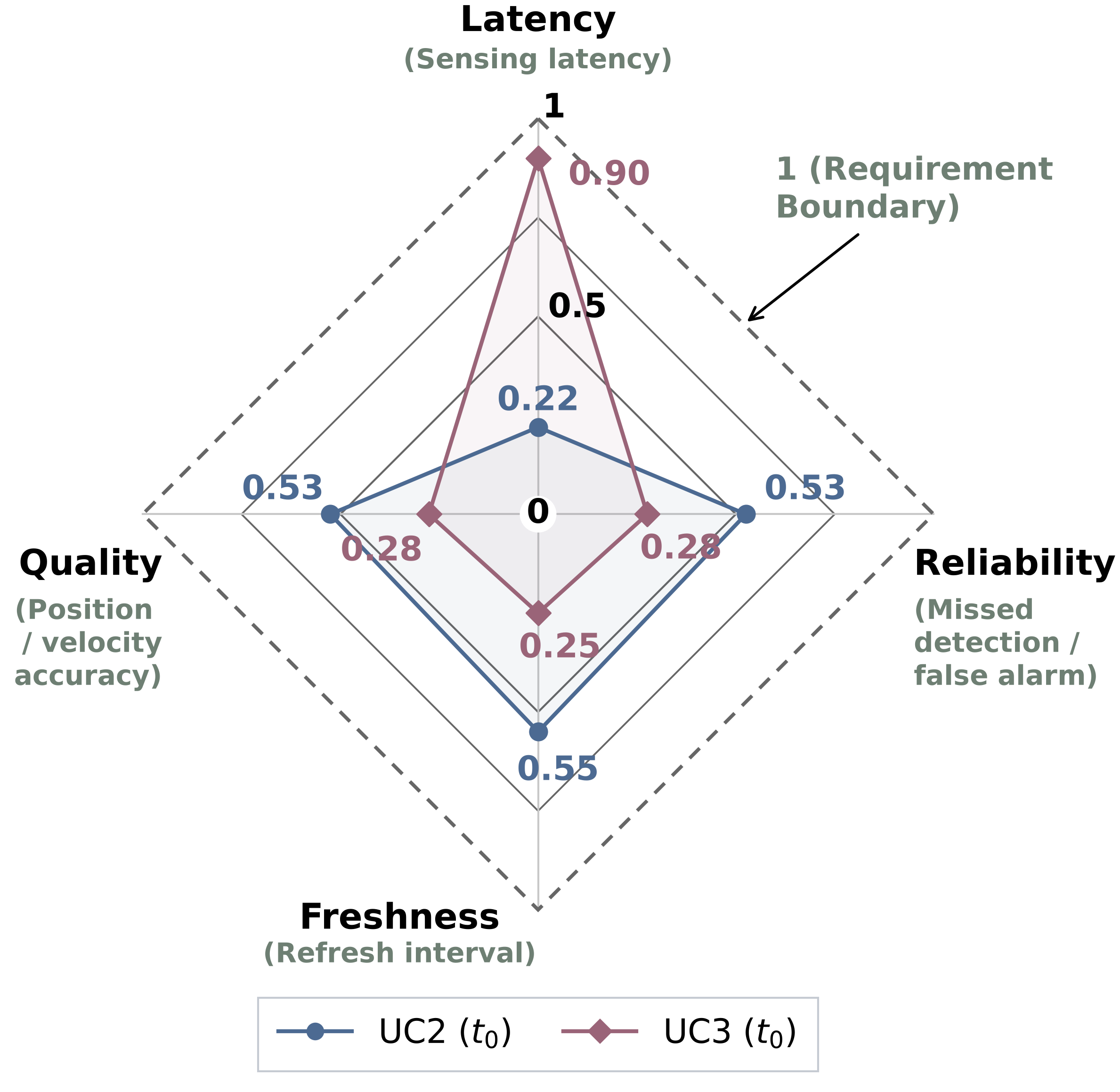}
    }
    \hfill
    \subfigure[]{
        \includegraphics[
            width=0.31\textwidth
        ]{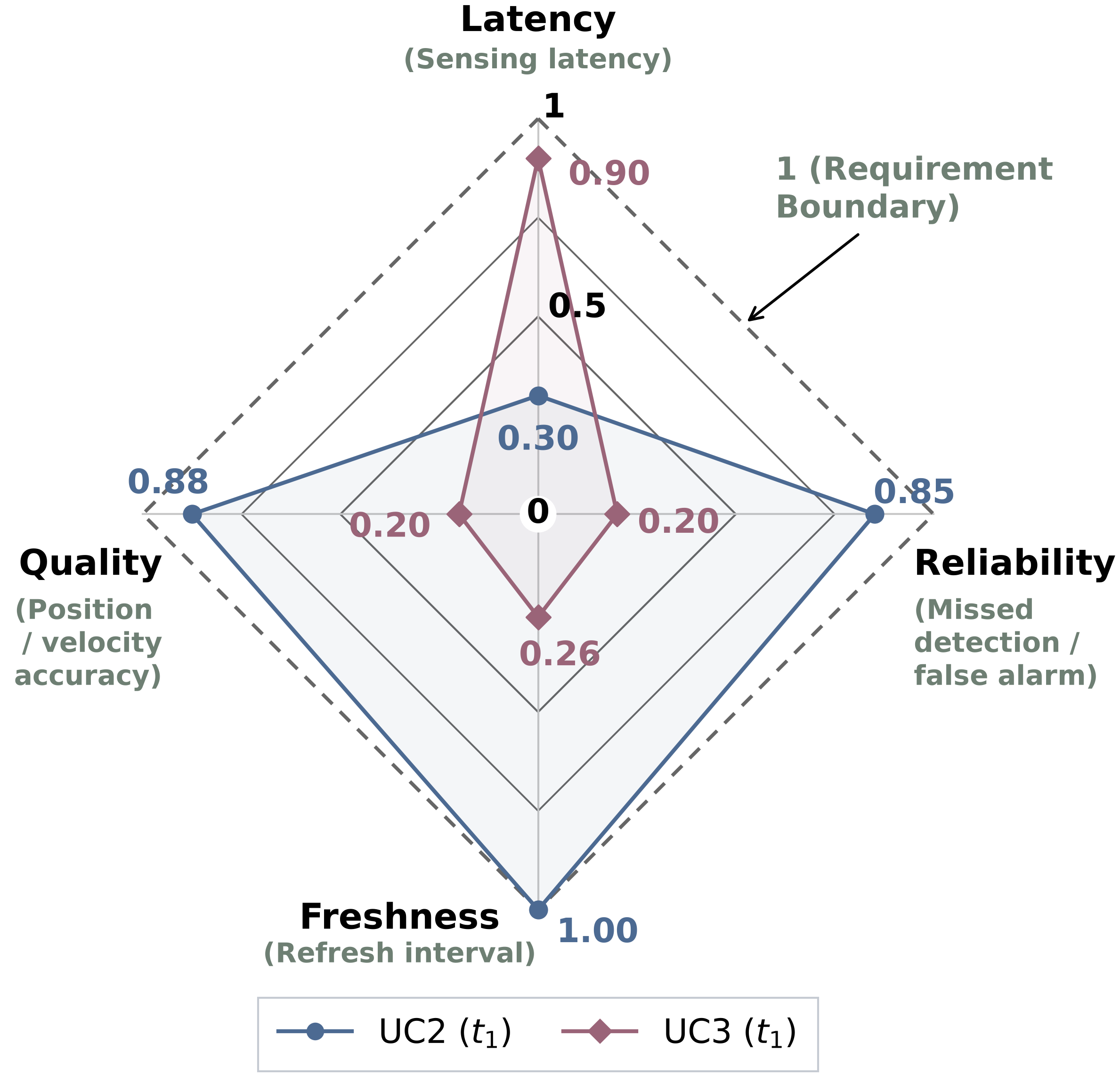}
    }
    \caption{
        Normalized performance comparison of feasible sensing configurations:
        (a) factory robot collision avoidance;
        (b) UAV flight trajectory tracking at $t_0$;
        (c) UAV flight trajectory tracking at $t_1$.
    }
    \label{fig:case_study}
\end{figure*}

Table~\ref{tab:case-results}(a) compares the illustrative performance of the four candidate configurations across the applicable QoS dimensions. $RC_1$ provides the lowest latency but violates multiple sensing quality requirements, while $RC_4$ provides the highest sensing quality but violates the latency requirement. Thus, both are excluded, leaving $RC_2$ and $RC_3$ as feasible candidates. $RC_2$ has lower latency, whereas $RC_3$ provides better reliability, freshness, and sensing-quality performance. If the selection were based only on sensing latency, $RC_2$ would be preferred over $RC_3$. However, SESO-ISAC evaluates the feasible candidates across all applicable QoS dimensions of the requested sensing service. Figure~\ref{fig:case_study}(a) compares the feasible candidates after normalizing each performance metric and aggregating them within the corresponding QoS dimensions. The normalized Euclidean distances are $0.734$ for $RC_2$ and $0.503$ for $RC_3$, so SESO-ISAC selects $RC_3$. This case illustrates that a configuration favored by a single QoS dimension is not necessarily preferred when complete end-to-end sensing configurations are evaluated across multiple applicable QoS dimensions.

\subsection{Case 2: UAV flight trajectory tracking}
We consider a sensing service in which a network-side SSC continuously tracks the position and velocity of a UAV moving along a predefined flight route. In this use case, the sensing operation may change as the UAV moves to maintain sensing service continuity~\cite{TR22.837}. SESO-ISAC considers three candidate configurations: 1) $UC_1$ uses RAN mono-static sensing with RAN-local processing; 2) $UC_2$ uses RAN--RAN bi-static sensing with local processing at the receiving RAN; and 3) $UC_3$ combines observations from multiple SenEs through network-side processing. The reporting path is RAN-to-network for $UC_1$ and $UC_2$ and network-local for $UC_3$.

At the initial time $t_0$, Table~\ref{tab:case-results}(b) shows that $UC_1$ is infeasible, leaving $UC_2$ and $UC_3$ as feasible candidates. $UC_2$ has lower latency, whereas $UC_3$ provides better reliability, freshness, and sensing-quality performance. Figure~\ref{fig:case_study}(b) compares the feasible candidates after normalizing each performance metric and aggregating them within the corresponding QoS dimensions. The resulting normalized distances are $0.475$ for $UC_2$ and $0.506$ for $UC_3$. Accordingly, $UC_2$ is selected under the current sensing execution context. As the UAV moves, changes in sensing geometry and SenE availability may require re-evaluation of SenE participation, Tx/Rx role assignments, and sensing modes.

At time $t_1$, SESO-ISAC re-evaluates the candidates under the changed sensing execution context. $UC_2$ remains feasible, but its normalized distance increases to $0.803$, while that of $UC_3$ decreases to $0.489$. As shown in Fig.~\ref{fig:case_study}(c), $UC_3$ becomes the preferred candidate, and SESO-ISAC changes the configuration from $UC_2$ to $UC_3$. This case illustrates that a context change can alter the relative performance of feasible configurations, triggering re-evaluation and configuration reselection when another candidate becomes preferred.

\section{Open Research Issues}
\label{Sec:issues}
In this section, we discuss open research issues for service-aware sensing operation orchestration in 6G ISAC.

\subsection{AI Agent and Data Framework Integration}
The 6G sensing lifecycle may involve the SenF, an artificial intelligence (AI) agent, and the 6G data framework. The AI agent may interpret intent-based sensing requests and coordinate sensing with other 6G services, while the data framework may support sensing-data and result management. When these functions jointly support a sensing service, orchestration responsibility and coordination of service states and QoS remain unclear. Future studies should investigate integrated orchestration mechanisms across the SenF, AI agent, and data framework.

\subsection{Heterogeneous Sensing Data Integration}
6G ISAC may use sensing data generated by heterogeneous SenEs and potentially non-3GPP sensing sources. Such sensing data may differ in measurement characteristics, timing, data representation, and quality. Although SESO-ISAC can compose end-to-end configurations involving multiple SenEs, the use of heterogeneous sensing data introduces additional requirements for joint processing and fusion. Future studies should investigate mechanisms for representing, aligning, and combining heterogeneous sensing data across different sensing sources.

\subsection{Security and Privacy for Sensing Services} 
Sensing services may involve information about the locations, movements, and behaviors of people and objects, necessitating security and privacy protection throughout the sensing lifecycle. Service authorization, UE authorization, and protection of sensing-data paths may depend on the selected sensing configuration and processing point. Future extensions of SESO-ISAC should therefore incorporate authorization and privacy policies as constraints for candidate generation and feasibility evaluation.

\section{Conclusion}
\label{Sec:Conclusion}
This article proposed SESO-ISAC for service-aware sensing operation orchestration in 6G ISAC. SESO-ISAC composes, evaluates, and selects complete end-to-end sensing configurations based on sensing service requirements and the sensing execution context, and re-evaluates candidate configurations as the context changes. Through two case studies, we illustrated multi-dimensional configuration selection and mobility-driven re-evaluation. Future work will investigate adaptive configuration selection that accounts for heterogeneous sensing data, service-specific QoS priorities, and uncertainty in candidate performance.

\begin{IEEEbiographynophoto}{YOUBIN JEON} received the B.S. degree from Myongji University, Korea, in 2015, and the Ph.D. degree from Korea University, Korea, in 2025. In 2019, she worked as a software engineer at Shinhan DS. She is currently with the 6G Communication Standard Task at LG Electronics. Her research interests include 5G/6G integrated sensing and communication (ISAC), mobile core networks, network automation, and AI-enabled networking. Contact her at \href{mailto:youbin.jeon@lge.com}{youbin.jeon@lge.com}.
\end{IEEEbiographynophoto}

\begin{IEEEbiographynophoto}{LAEYOUNG KIM} received the Ph.D. in computer science from Yonsei University, Seoul, Republic of Korea, and is working for LG Electronics. Her work focuses on system architecture standardization, mainly covered by 3GPP SA WG2 and SA. Contact her at \href{mailto:laeyoung.kim@lge.com}{laeyoung.kim@lge.com.}
\end{IEEEbiographynophoto}

\begin{IEEEbiographynophoto}{MYUNGJUNE YOUN} received the Ph.D. in Electrical and Electonic Engineering from Yonsei University, Seoul, Republic of Korea, and is working for LG Electronics. His work focuses on system architecture standardization, mainly covered by 3GPP SA WG2. Contact him at \href{mailto:m.youn@lge.com}{m.youn@lge.com.}
\end{IEEEbiographynophoto}

\begin{IEEEbiographynophoto}{SANGHEON PACK} received the B.S. and Ph.D. degrees from Seoul National University, Korea, in 2000 and 2005, respectively. In 2007, he joined the faculty of Korea University, Korea. He is currently a professor at the School of Electrical Engineering. His research interests include network softwarization and mobile edge computing. Contact him at \href{mailto:shpack@korea.ac.kr}{shpack@korea.ac.kr}.
\end{IEEEbiographynophoto}

\end{document}